\documentclass[prl,aps,twocolumn]{revtex4}
\usepackage{graphicx} 
\usepackage[usenames]{color}
\usepackage{amsmath,amssymb}
\usepackage{gensymb}
\usepackage{natbib}
\usepackage{xcolor}
\def\strutdepth{\dp\strutbox}
\def\nw#1{\strut\vadjust{\kern-\strutdepth\vtop to0pt{\vss\hbox to\hsize
{\hskip\hsize\hskip5pt$\leftarrow$\hss\strut}}}{\em #1}}
\usepackage{rotating}

\pdfoutput=1

\begin{document}

\title{Dynamical mechanism for non-locality in dense granular flows}
\author{{Mehdi Bouzid}}
\author{Martin Trulsson}
\author{Philippe Claudin}
\author{Eric Cl\'ement}
\author{Bruno Andreotti}
\affiliation{Physique et M\'ecanique des Milieux H\'et\'erog\`enes, PMMH UMR 7636 ESPCI -- CNRS -- Univ.~Paris-Diderot -- Univ.~P.M.~Curie, 10 rue Vauquelin, 75005 Paris, France}
%\email{mehdi.bouzid@espci.fr}
 
%___________________________________________________________________________
\begin{abstract}
The dynamical mechanism at the origin of the non-local rheology of dense granular flows is investigated trough discrete element simulations. We show that the influence of a shear band on the mechanical behavior of a distant zone is contained in the spatial variations observed in the network of granular contacts. Using a micro-rheology technique, we establish that the exponential responses hence obtained, do not proof the validity of a mechanical activation process as previously suggested, but stem from the spatial relaxation of the shear rate as a direct consequence of a macroscopic non-local constitutive relation. Finally, by direct visualization of the local relaxation processes, we dismiss the kinetic elasto-plastic picture, where a flow is conceived as a quasi-static sequence of localized plastic events interacting through the stress field. We therefore conclude in favor of the jamming scenario, where geometrical constrains lead to coherent non-affine displacements along floppy modes, inherently non-local.
\end{abstract}

\date{\today}

\maketitle

%___________________________________________________________________________
Soft amorphous materials are disordered assemblies of interacting particles that can resist shear like a solid, but which flow like a liquid under a sufficiently large applied shear stress, see \cite{CPB09} and related references. The solid-like state is usually associated with highly multi-stable energy landscapes, whose origin allows one to classify these materials: entropy for colloidal suspensions \cite{HW12}, free energy for foams or emulsions (surface tension) as well as for soft elastomeric particles (elasticity) \cite{CBML03}, and geometry for a granular material (volume times imposed pressure) \cite{AFP13}. The rigidity transition is accordingly named colloidal glass transition, elasto-plastic depinning transition and jamming transition for these three classes of systems. Like standard phase transitions, these dynamical transitions are associated with diverging length scales quantifying the degree of cooperativity in the flow or in the deformation response. Amongst the consequences, experiments have shown that the rheology of granular materials and emulsions is non-local: the stress at a given location does not depend only on the local strain rate but also on the degree of mobility in the surrounding region \cite{GCOAB08,NZBWvH10,RFP11,WH14}. Non-local constitutive relations were recently proposed and successfully tested against numerical and experimental results \cite{BCA09,PF09,BTCCA13,HK13}. As these relations are derived from a gradient expansion in the order parameter, analogous to a Ginzburg-Landau theory, they reflect the symmetries, but not the nature of the underlying dynamical mechanisms.

%%%%%%%%%%%%%%%
\begin{figure}[t!]
\includegraphics{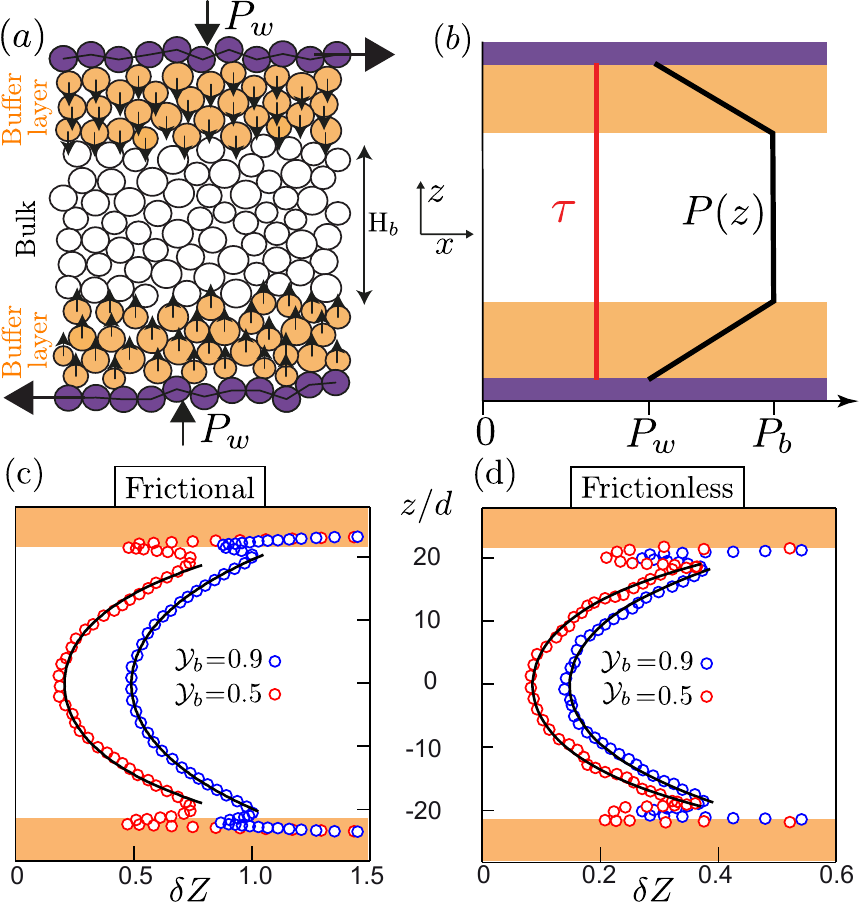}
\vspace{- 4 mm}
\caption{(Color online)
(a-b) Schematics of the numerical shear cell. The gravity-like forces applied to the grains in the buffer layers located between the walls and the bulk (a) enable the control of the pressure profile $P$ (black line) and of the shear stress profile $\tau$ (red line) across the cell (b), see \cite{BTCCA13}. (c-d)  Profiles of the distance to isostaticity $\delta Z(z)$ at two values of the yield parameter $\mathcal{Y}_b$ (see legends), for frictionless (c) and for frictional (d) grains. Symbols: numerical data. Solid lines: $A\cosh(z/\xi)$ where $A$ and $\xi$ are fitting parameters.}
\vspace{- 5 mm}
\label{Fig1}
\end{figure}
%%%%%%%%%%%%%%%

%%%%%%%%%%%%%%%
\begin{figure}[t!]
\includegraphics{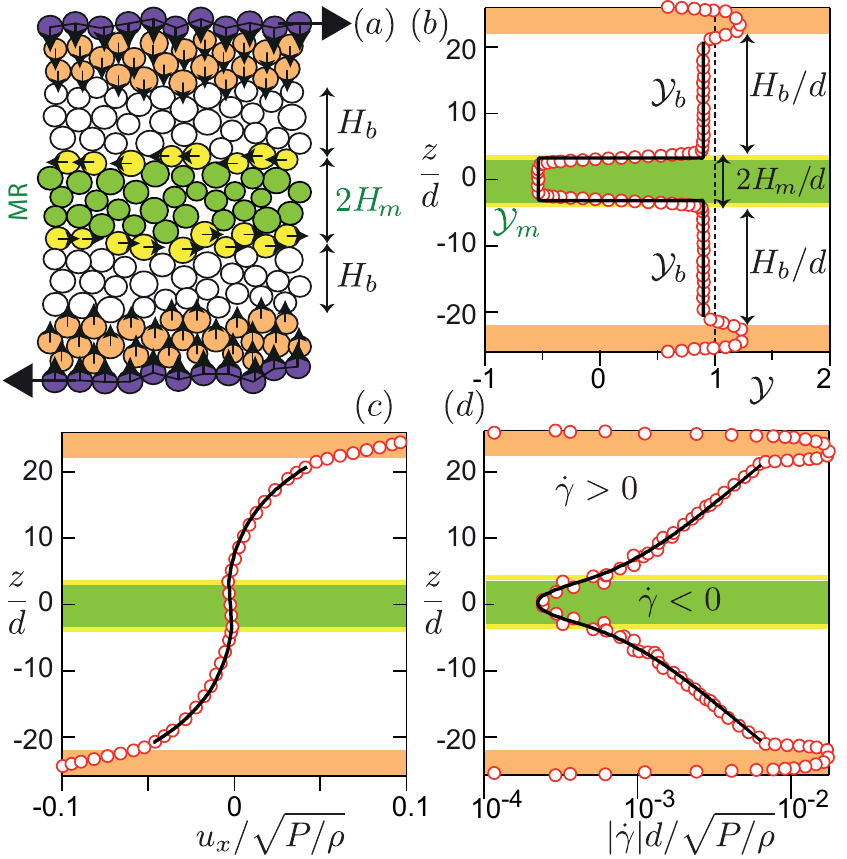}
\vspace{- 5 mm}
\caption{(Color online)
(a) Schematic of the numerical set-up with a micro-rheometer in the center (in green in all panels). (b) Profile of the yield parameter $\mathcal{Y} = \tau/(\mu P)$, with here $\mathcal{Y}_b =0.9$ and $\mathcal{Y}_m =-0.5$. Corresponding velocity (c) and shear rate (d) rescaled profiles. Symbols: numerical data. Solid lines: predictions of the non-local rheology without any adjustable parameter.}
\vspace{- 5 mm}
\label{Fig2}
\end{figure}
%%%%%%%%%%%%%%%

The aim of this Letter is precisely to investigate the dynamical mechanisms behind non-locality in dense granular materials. Three scenarios have emerged so far to explain non-locality. (i) For hard non-deformable grains, the topological analysis of the contact network provides a state parameter, the mean contact number $Z$, to distinguish between solid-like and liquid-like states. Close to the jamming transition, flow is only possible along floppy modes, by essence spread in space and prescribing the cooperative motion of the particles \cite{ABH12}. The evolution of the contact network results from the rapid formation of force chains followed by a slow zig-zag instability of these structures \cite{LDW12}. (ii) For very elastic particles e.g. for foams, emulsions or glassy Lennard Jones phases, the dynamics in the quasi-static regime is controlled by elasto-plastic events~\cite{TLB06,LC09,BCA09,LP09}: when sheared, energy is slowly stored and rapidly released through scale-free avalanches, in close analogy with the depinning transition of an elastic line. Recent experiments have suggested that an assembly of rigid particles could similarly present an intermittent succession of plastic events localized in space and time \cite{ANBCC12}. (iii) The third picture is based on an analogy with Eyring's transition state theory for the viscosity of liquids \cite{PF09}, where mechanical fluctuations would play the role of temperature in thermal systems. Each plastic rearrangement  is assumed to generate at random a new realization of the forces on the contact network, allowing for the formation of new weak zones where the next rearrangement will occur. In this letter, we will test successively these three pictures using discrete element simulations.

\emph{The numerical system} is a 2D shear flow, similar to that used in~\cite{BTCCA13}, and composed of grains confined between two rough solid walls. By means of gravity-like vertical forces, the normal stress $P$ is increased gradually in two thin boundary layers located close to the walls (Fig.~\ref{Fig1}a) and reaches a constant value $P_b$ in the bulk (Fig.~\ref{Fig1}b). The yield parameter $\mathcal{Y}=\tau/(\mu_c P_b)$ based on the shear stress $\tau$, which is kept homogeneous across the cell, and on the dynamical friction coefficient $\mu_c$ is the central control parameter (Fig.~\ref{Fig1}b). In order to emphasize non-local effects, we focus here on the case $\mathcal{Y}<1$, were the grains flow \emph{below the yield stress} under the influence of the distant shear zone.

\emph{The jamming framework~}provides a picture of non-locality essentially based on the contact network geometry. The central criterion to render the status of a packing of hard grains is the mean contact number per grain $Z$. Following Maxwell rigidity criterion, we define the distance $\delta Z$ to isostaticity as the difference between the number of constrains and the number of degrees of freedom (the number of force components). One must distinguish the case were a Coulomb friction between the particles, of coefficient $\mu$, is introduced from the frictionless case ($\mu=0$):
\begin{equation}
\delta Z = \left\{
  \begin{array}{lr}
    2D-Z & \rm{when}~ \mu=0\\
    D+1+\zeta-Z &  \rm{when}~ \mu\neq0.
  \end{array}
\right.
\end{equation}
were $D$ is the space dimension and $\zeta$ the fraction of sliding contacts \cite{MVH}.  A system is ''solid'', i.e. rigid, when hyperstatic ($\delta Z<0$) and ''liquid'' when hypostatic ($\delta Z>0$). Importantly, this criterion is entirely non-local because it is global, which means that liquid or solid regions can hardly be isolated.

In our numerical system, $Z$ and $\delta Z$ are defined as Eulerian fields, by averaging over realizations or, equivalently, as we work with simulations in a statistically steady state, over time. In Fig.~\ref{Fig1}c and Fig.~\ref{Fig1}d , we display profiles of $\delta Z$, both in the frictional and frictionless cases. We first notice that all values are positive. These profiles can be fitted to a shape of the form $A \cosh(z/\xi)$. This means that $\delta Z$  spatially relaxes toward $0$ from both walls, in an exponential manner. The whole system (the boundary layers and the bulk) constitutes a single continuous liquid phase. From the jamming perspective, the flow in the bulk, observed when the stress is lower than the yield stress,  results from a lack of coordination in the contact network. Particles are expected to move cooperatively along floppy modes whose characteristic size diverges at the jamming point. A flowing region therefore favors the motion around it. As it affects the contact network geometry in its neighborhood, one expects the corresponding non-local correction to the rheology to remain finite in the limit of vanishing shear rate.

Since there is a relation between $Z$ and the rescaled shear rate $I=|\dot \gamma| d / \sqrt{P/\rho}$~\cite{cDEPRC05}, this jamming picture is consistent with the non-local constitutive relation based on a gradient expansion in $I$, derived, calibrated and tested in a previous letter \cite{BTCCA13}:
\begin{eqnarray}
|\mathcal{Y}|=\frac{|\tau|}{\mu_c P}=\frac{\mu(I)}{\mu_c} \left[1-\chi(\kappa)\right]
\quad{\rm where}\quad
\kappa = d^2\frac{\nabla^2 I}{I},
\label{rheononlocalI}
\end{eqnarray}
where $\mu(I)$ is the effective friction coefficient for homogeneous shear flows. The Laplacian $\nabla^2 I$ encodes the non-local effect, assumed isotropic; it is rescaled by $I$ to form $\kappa$, which remains finite as $I \to 0$. The linear expansion of the non-local correction is written as $\chi(\kappa)\simeq \nu \kappa+\mathcal{O}(\kappa^2)$, with  $\nu \simeq 8$. 

\emph{Mechanical activation {\it \`a la Eyring}~}has recently been put forward in the interpretation of the experiments of  Reddy~\textit{et al} \cite{RFP11}. The conceptual idea of these authors was to measure the rheology within a shear band that is below the yield condition, but under the remote influence of a shear zone forced by rigid boundaries. As discussed below, an exponential behavior of the local shear rate with the distance to the yield condition has been interpreted as a Boltzmann-like factor, with an effective `temperature' that would result from mechanical activation. 

To investigate whether this behavior is consistent with the previous picture, we probe numerically the rheological response with a local micro-rheometer in the spirit of the set-up proposed by \cite{RFP11} (Fig.~\ref{Fig2}a and \cite{OSM}).  Bulk horizontal forces acting on grains are added to impose a different shear stress in a small central region. Fig.~\ref{Fig2}b shows the resulting profile of the yield parameter $\mathcal{Y}$.  As before, the flow is driven by the two boundary layers located close to the moving walls, which are above the yield condition. In the bulk, the flow is kept below the yield condition (i.e. $\mathcal{Y}_b<1$). The shear rate at the locations where $\mathcal{Y}$ crosses $1$ is noted $\dot \gamma_{b}$ and is considered as a boundary condition for the bulk rheology. The micro-rheometer is the central region where a different yield parameter $\mathcal{Y}_{m}$ is imposed. Its thickness is noted $2H_{m}$ and its location is $H_b$ away from the boundaries (Fig.~\ref{Fig2}a). The relation between the shear rate $\dot \gamma_{m}$ in the center of the micro-rheometer and the rescaled shear stress $\mathcal{Y}_{m}$  gives access to the local rheological properties. In practice, $\dot \gamma_{m}$ depends on four other parameters: $\dot \gamma_{b}$, $\mathcal{Y}_b$, $H_b$ and $H_{m}$. In the experiments of  \cite{RFP11}, the micro-rheometer was a small rod immersed in a granular medium confined in a Couette cell and submitted to an external force, which, once rescaled by its critical value, plays the role of $\mathcal{Y}_{m}$. The rod creep velocity is the analogous of $\dot \gamma_{m}$ and the shear rate $\dot \gamma_{b}$  is the analogous of the rotation velocity of the inner cylinder. Note that contrarily to the numerical set-up, the radial profile of the yield parameter $\mathcal{Y}$ is not homogeneous in the Couette cell. The key observations reported in \cite{RFP11}  are recovered in the numerics. (i) The shear rate $\dot \gamma_{m}$ in the micro-rheometer is proportional to the shear rate $\dot \gamma_{b}$ imposed by the boundary (Fig.~\ref{Fig2}a). (ii)  $\dot \gamma_{m}$ (roughly) decreases exponentially with the distance to the yield condition, measured by $1-|\mathcal{Y}_{m}|$ (Fig.~\ref{Fig2}b).

This exponential dependence can be recovered by the non-local rheology: linearizing the constitutive relation (\ref{rheononlocalI}) around the critical state ($\dot \gamma=0$), the inertial number $I$ decays exponentially over a relaxation length $\ell(\mathcal{Y}) \sim d \sqrt{\nu}\;||\mathcal{Y}|-1|^{-1/2}$ \cite{BTCCA13,OSM}. Using the continuity of $I$ and $dI/dz$ at the interface $z=H_m$, it is straightforward to integrate Eq.~\ref{rheononlocalI} \cite{OSM}. In the limit $H_b\gg\ell(\mathcal{Y}_b)$, the solution reads:
\begin{equation}
\frac{{\dot \gamma}_{b}}{|{\dot \gamma}_{m}|} = \frac{1}{2}
e^{\frac{H_b}{\ell(\mathcal{Y}_b)}} \!
\left[\cosh\!\left( \frac{H_{m}}{\ell(\mathcal{Y}_{m})}\right) \! +  \frac{\ell(\mathcal{Y}_b)}{\ell(\mathcal{Y}_{m})}\sinh\!\left( \frac{H_{m}}{\ell(\mathcal{Y}_{m})}\right) \! \right] \!. \qquad
\label{gammadots}
\end{equation}
This solution is displayed in (Fig.~\ref{Fig2}) and is in excellent agreement with the numerical simulation results, without any adjustable parameter, once the constitutive relation (\ref{rheononlocalI}) has been calibrated in the simple shear case. In particular, the two key properties (i) and (ii) are recovered. In conclusion, the exponential behaviors found either experimentally~\cite{RFP11} or in the present simulations result from a linear relaxation of the shear rate. They do not point to a specific mechanism like mechanical activation, but to any mechanism that, like jamming, is roughly isotropic and leads to a finite non-local correction at vanishing shear rate.

%%%%%%%%%%%%%%%%
\begin{figure}[t!]
\includegraphics{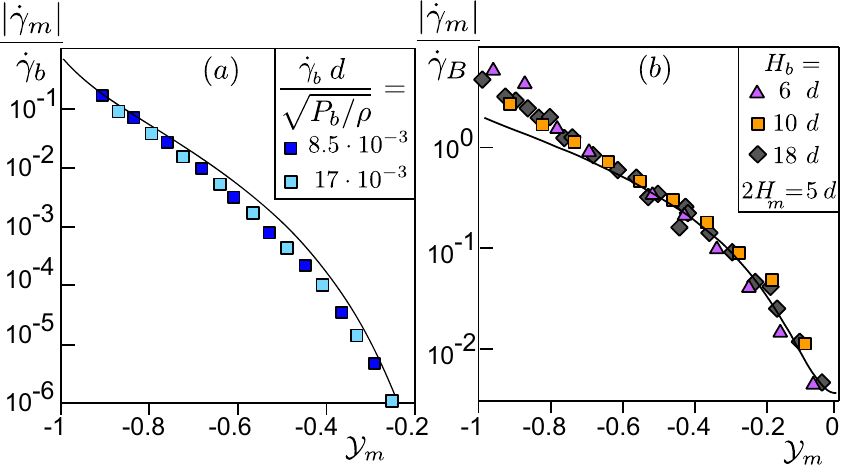}
\vspace{- 4 mm}
\caption{(Color online) Shear rate $\dot \gamma_{m}$ as a function of the yield parameter $\mathcal{Y}_{m}$ in the micro-rheometer. (a) Evidence for linearity of the response to the shear rate in the boundary layers $\dot \gamma_{b}$. (b) Influence of the distance $H_b$: data collapse once rescaled by the factor $\dot \gamma_B \equiv \dot \gamma_{b}\;\exp\left[-H_b/\ell(\mathcal{Y}_b)\right]$, where $\ell$ is the relaxation length of the non-local rheology \cite{BTCCA13}. Solid lines: predictions of Eq. (\ref{rheononlocalI}) without any adjustable parameter.}
\vspace{- 5 mm}
\label{Fig3}
\end{figure}
%%%%%%%%%%%%%%%

%%%%%%%%%%%%%%%
\begin{figure}[t!]
\includegraphics{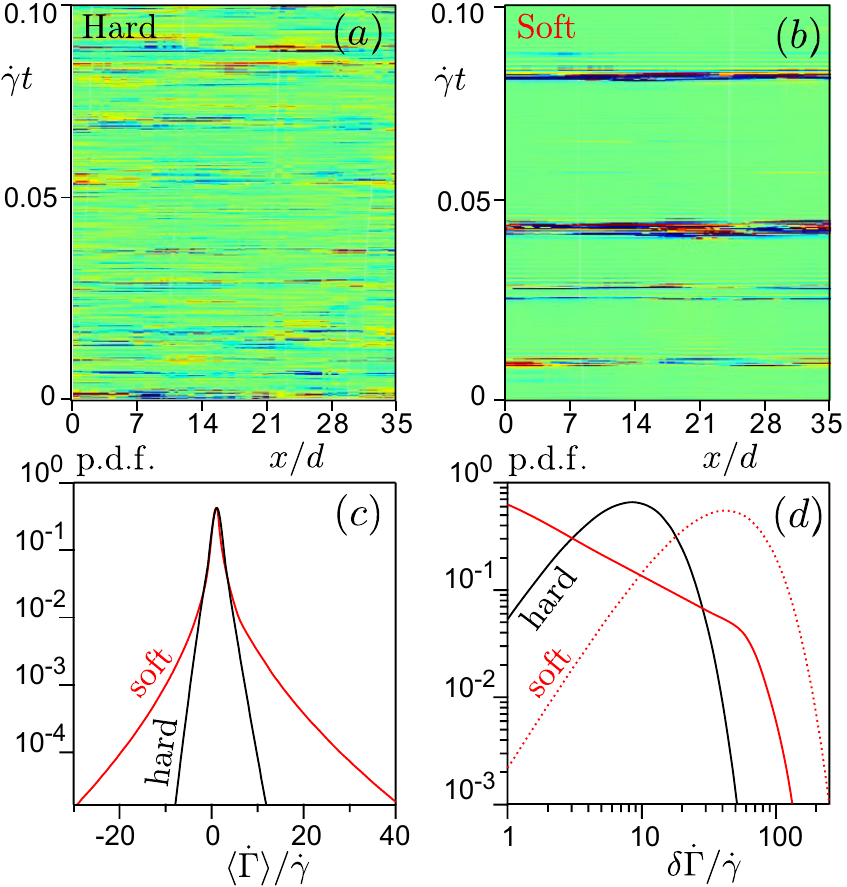}
\vspace{- 4 mm}
\caption{(Color online) Space-time diagrams showing the local contribution $\dot \Gamma$ to the shear rate $\dot \gamma$, measured on the central line of the cell, for a system of hard (a) and soft (b) grains. Color code from blue ($\dot \Gamma = -40 \dot \gamma $) to red ($\dot \Gamma = 40 \dot \gamma $). (c) Probability distribution function (PDF) over time of the space average $\langle \dot \Gamma \rangle$, for hard (black line) and soft (red line) grains. (d)  PDF of the spatial standard deviation $\delta \dot \Gamma$. Red dotted line: $\delta \dot \Gamma$ computed when $|\langle \dot \Gamma \rangle| \geq 5 \dot \gamma$ (soft system).}
\vspace{- 5 mm}
\label{Fig4}
\end{figure}
%%%%%%%%%%%%%%

\emph{The elasto-plastic~}picture assumes that the material behaves most of the time like a solid, but presents local and short lived plastic events, in regions sometimes called `shear transformation zones' \cite{FL98}. The associated scenario is a localized rupture initiation followed by a scale free avalanche of localized events. In order to investigate whether this picture constitutes an alternative to the jamming scenario to interpret the non-local nature of the granular rheology, we compared the dynamics of two otherwise identical systems composed of hard and soft grains. The presence of localized plastic events is usually based on a visual inspection of different fields. The squared deviation from an affine deformation on a local scale has for instance been proposed as a field indicating plastic activity \cite{FL98,LC09}. However, such a quantity, as well as all those based on the non-affine velocity field, characterize fluctuations around the mean flow, and not the local contribution of a certain area to the mean flow. We wish here to propose a practical definition of these events, based on the quantitative criterion that they must be separated in time and localized in space. Importantly, to match their role played in elasto-plastic models \cite{FL98,LC09,BCA09,HK13}, they must also contribute additively to the average shear rate $\dot \gamma$.

In order to detect localized plastic events, we have built a coarse-grained field $\dot \Gamma(\vec r,t)$ reflecting, at time $t$, the local contribution to $\dot \gamma$ of a small region around the position $\vec r$. We impose that the time average of $\dot \Gamma$ must everywhere give $\dot \gamma$. A coarse-graining method similar to that proposed for the stress tensor \cite{GG01,GG02} is adapted here to the computation of velocity differences -- see \cite{OSM} for the explicit expression of $\dot \Gamma$. We display in Fig.~\ref{Fig4} for a system of rigid (a) and soft grains (b), spatio-temporal diagrams showing the local contribution $\dot \Gamma$ to the shear rate $\dot \gamma$, measured on the central line of the cell (without micro-rheometer, i.e. in the system schematized in Fig.~\ref{Fig1}a). We observe contrasted behaviors in the two cases. In the soft system, nothing much happens most of the time, except for short periods of intense activity, associated with a cascade of plastic events. Conversely, the hard system presents more moderate but permanent fluctuations even for asymptotically small $\dot \gamma$.

To make these observations quantitative, Fig.~\ref{Fig4}c shows the probability distribution function (PDF) over time of $\langle \dot \Gamma\rangle$, which is the space average of $\dot \Gamma$ over the cell. In panel (c), we similarly display the PDF of the spatial standard deviation $\delta \dot \Gamma$. The hard-particle system presents a narrow gaussian distribution of $\langle \dot \Gamma\rangle$ around $\dot \gamma$, while the PDF corresponding to the soft system shows stretched tails, which are due to a very intermittent behavior associated with these plastic events (Fig.~\ref{Fig4}c). The PDF of $\delta \dot \Gamma$ provides informations about spatial heterogeneities in the system. The peak of the black line around $10 \dot \gamma$ in Fig.~\ref{Fig4}d indicates that they are large and permanent in the hard system. For the soft system, the PDF shows an algebraic decay, which means that the field $\dot \gamma$ is homogeneous most of the time. However, when the computation of $\delta \dot \Gamma$ is restricted to the periods of time where plastic events occur (periods where $|\langle \dot \Gamma \rangle|$ is larger than $5 \dot \gamma$ in Fig.~\ref{Fig4}d), its PDF also presents a peak: in the soft system, plastic events are associated with a very heterogeneous field of $\dot \Gamma$. In conclusion, an assembly of rigid particles does not present local plastic events when sheared permanently. Its dynamics is not intermittent but presents spatial heterogeneities.

%___________________________________________________________________________
\emph{Conclusion}~--~The study reported in this Letter allows us to conclude that the non-locality in dense granular flows of rigid particles is consistent with the existence of floppy modes characterizing the jamming picture. The most striking observation is the lack of localized plastic events in such \emph{liquid} flows. This seems to be in contradiction with the fact that plastic events have been observed experimentally in both soft \cite{KD03,GCOAB08} and hard \cite{ANBCC12,LBetal14} particle systems. However, from the jamming point of view, all these systems are \emph{solid}: foams and emulsions present plastic events when they flow at high volume fraction; a granular sample that is not prepared in the critical state by shearing is generically hyper-static, due to friction. This suggests that $\delta Z$ (or the volume fraction) could be the parameter controlling the cross-over between a non-local behavior ruled by the contact network geometry and another controlled by local plastic events. Further numerical studies are required to investigate the nature of this cross-over.
 
Moreover, we have shown that the non-local constitutive relation proposed by Bouzid et al. \cite{BTCCA13} is able, without any other adjusting parameter or any assumption on the mechanisms at work, to predict the inhomogeneous fields stemming from a micro-rheological response. This results gives further incentive to clarify the relation between the constitutive coefficients and the most fundamental aspects of the soft-modes dynamics in the vicinity of the jamming transition.

%___________________________________________________________________________
BA is supported by Institut Universitaire de France. This work is funded by the ANR JamVibe and a CNES grant.

%%%%%%%%%%%%%%%%%
\appendix

\vspace*{1cm}

{\large \bfseries \centerline{Supplementary Material}}

%----------------------------------------------------------------------------------------------------------------
\section{Numerical simulations}

\begin{figure}[t!]
\includegraphics{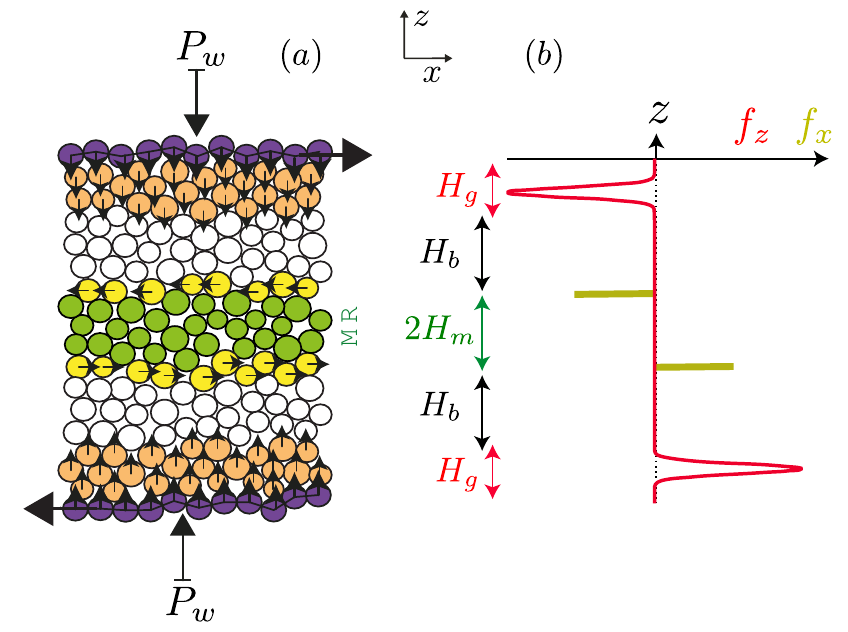}
%\vspace{-5 mm}
\caption{a) Schematics of the numerical set-up. Walls are depicted by dark purple circles. $P_w$ is the external wall pressure. (b) $z$-profiles of the gravity-like forces $f_x$ (yellow) and $f_z$ (orange).}
%\vspace{-5 mm}
\label{SI1}
\end{figure}

We consider a two-dimensional system constituted of  $N \simeq 2\cdot 10^3$ circular particles of mass $m_i$ and diameter $d_i$, with a $\pm$20\% polydispersity. $i=1,N$ is the particle label. The grains are confined in a shear cell composed of two rough walls moving along the $x$-direction at opposite constant velocities, see Fig.~\ref{SI1}a. These walls are made of similar grains, but glued together. We call $2H \simeq 55 d$ the distance between the walls. Their position is controlled to ensure a constant normal stress $P_w$ at the walls -- the distance $2H$ then fluctuates during the simulations (typically by a fraction of the grain diameter). Periodic boundary conditions are applied along the $x$-direction. The particle and wall dynamics are integrated using the Verlet algorithm. Contact forces between particles are modeled as viscoelastic forces, with a possibility to add a Coulomb friction along the tangential direction \cite{Cundall79,cDEPRC05,Luding06}. The normal spring constant $k_n$ is chosen sufficiently large (i.e. $k_n/P>10^{4}$) to be in the rigid asymptotic regime where the results are insensitive to its value. The tangential spring constant is $k_t=0.5 k_n$. The Coulomb friction coefficient is chosen equal to $\mu_p=0.4$ and set to $0$ for frictionless grains. Damping parameters are chosen such that the restitution coefficient is $e \simeq 0.1$. We have checked that the results are independent of the value of $e$ in the range $0 \le e \le 0.9$.

%%%%%%%%%%
%%%%%%%%%%

\begin{figure}[t!]
\includegraphics{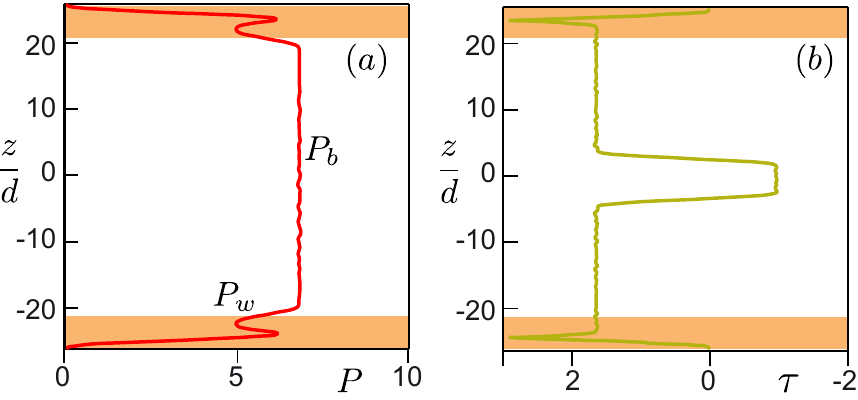}
%\vspace{-5 mm}
\caption {Typical profiles of the pressure $P$ (a) and of the shear stress $\tau$ (b).}
%\vspace{-5 mm}
\label{SI2}
\end{figure}

The particles in the regions next to the walls (orange grains in Fig.~\ref{SI1}a.) are submitted to gravity-like bulk forces along the $z$-direction \cite{BTCCA13}, with a gaussian spatial distribution:
\begin{eqnarray}
f_z^i(z_i) & = & \hat{f}_z m_i \left\{
- \exp \left[ -\frac{(z_i - H + \frac{1}{2}H_g)^2}{2\sigma^2} \right]
\right. \nonumber \\
& + & \left. \exp \left[ -\frac{(z_i+H-\frac{1}{2}H_g)^2}{2\sigma^2} \right].
\right\}
\label{fzMSB}
\end{eqnarray}
The origin of the $z$-axis is chosen at the center of the cell. $H_{g} = 5d$ is the thickness of these buffer layers. The width of the distribution is one grain diameter: $\sigma = d$. $\hat{f}_z$ is the amplitude of the forcing. These forces are oriented downward at the top of the cell, and upward at the bottom, see Fig.~\ref{SI1}b. The pressure $P_b$ in the central region of the cell results from the external wall pressure $P_w$ and the sum of all these gravity-like forces (Fig.~\ref{SI2}a).

The secondary (stress controlled) rheometer (MR for `micro-theometer') is implemented by applying opposite horizontal forces on the grains located along the lines at $z=\pm H_m$:
\begin{equation}
f_x^i(z_i) =  \hat{f}_x m_i \left[ - \Pi \left( \frac{z_i - H_m}{w} \right) + \Pi \left( \frac{z_i+H_m}{w} \right) \right],
\label{fxSSR}
\end{equation}
where $\Pi$ is the rectangular function. Its effective width is the grain diameter: $w=d$.  $\hat{f}_x$ is the amplitude of the forcing. These forces are oriented leftward at the top of the MR, and rightward at the bottom, see Fig.~\ref{SI1}b. The resulting profile of the stress in the cell is given in Fig.~\ref{SI2}b. 

%%%%%%%%%%
%%%%%%%%%%

%----------------------------------------------------------------------------------------------------------------
\section{Non-local rheology}
\begin{figure}[t!]
\includegraphics [scale=1.5]{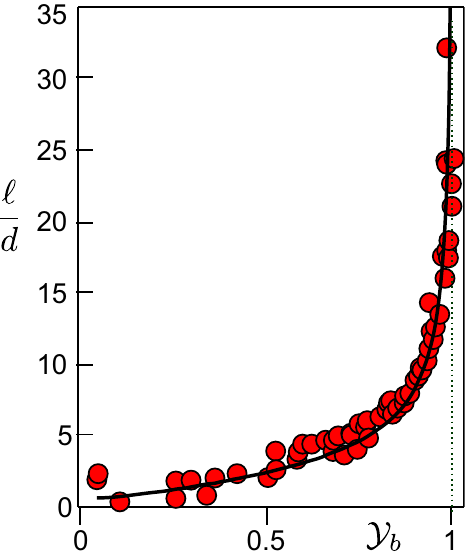}
%\vspace{-5 mm}
\caption{Relaxation length $\ell$ as a function of the yield parameter $\mathcal{Y}_b$ in the bulk. Symbols: numerical data. Solid line: fit of the data, with a diverging behavior scaling as $\sim |1-\mathcal{Y}_b|^{-1/2}$, see \cite{BTCCA13}.}
%\vspace{-5 mm}
\hspace{5mm}
\label{FigSI2}
\end{figure}

In \cite{BTCCA13}, we have proposed and calibrated a non-local constitutive equation for dense granular flows in terms of a gradient expansion of the inertial number $I=|\dot \gamma| d / \sqrt{P_b/\rho}$, given by:
\begin{eqnarray}
|\mathcal{Y}|=\frac{|\tau|}{\mu_c P}=\frac{\mu(I)}{\mu_c} \left[1-\chi(\kappa)\right],
\quad{\rm where}\quad
\kappa \equiv d^2\frac{\nabla^2 I}{I}.
\label{rheoNL}
\end{eqnarray}
The function $\chi$ is the non-local correction to the local rheology \cite{GDRMidi,cDEPRC05}. Consider the situation for which the base state is below the yielding condition $|\mathcal{Y}_b|<1$. Linearizing Eq.~(\ref{rheoNL}) around the critical state $I_b=0$ with $I=I_b+\delta I$, one gets $|\mathcal Y_b| = 1- \chi(\kappa)$, i.e.:
\begin{equation}
\chi^{-1}({1-|\mathcal Y_b|})= d^2\frac{ \nabla^2 (\delta I)}{\delta I}.
\label{chimoinsun}
\end{equation}
The solution of this second order differential equation in $\delta I$ is a combination of $\exp[\pm z/\ell(\mathcal Y_b)]$, with a relaxation length
\begin{equation}
\ell=\frac{d}{\sqrt{\chi^{-1} \left(1-|\mathcal{Y}_b|\right)}}.
\label{ellbelow}
\end{equation}
The measure of $\ell$ as a function of $\mathcal{Y}_b$ is displayed in Fig.~\ref{FigSI2}. Close to $|\mathcal{Y}_b|=1$, the non-local correction can be expanded as $\chi(\kappa) = \nu \kappa+\mathcal{O}(\kappa^2)$, where $\nu$ a numerical constant ($\nu \simeq 8$, see \cite{BTCCA13}). This implies that $\ell$ must diverge as $|1-\mathcal{Y}_b|^{-1/2}$ close to the yield condition. The phenomenological fit of the the numerical data (Fig.~\ref{FigSI2}) sets the calibration of the function $\chi(\kappa)$ via (\ref{ellbelow}), after which no more parameter of the rheological model is adjustable.

%%%%%%%%%%
%%%%%%%%%%

Let us now consider the situation were we add the micro-rheometer (MR) in the center of the cell, where a different yield parameter $\mathcal{Y}_{m}$ is imposed. The solution of the linearized constitutive relation takes, in the MR, the form
\begin{equation}
\dot\gamma=\dot\gamma_m\cosh[z/\ell(\mathcal{Y}_m)].
\label{gammadotinMR}
\end{equation}
Outside of the MR, it rather reads
\begin{equation}
\dot\gamma=\dot\gamma_+\exp[z/\ell(\mathcal{Y}_b)]+\dot\gamma_-\exp[-z/\ell(\mathcal{Y}_b)].
\label{gammadotoutMR}
\end{equation}
The constants $\dot\gamma_{\pm}$ are selected by the continuity of $I(z)$ and $dI/dz$ at the interface $z=H_m$. These conditions give:
\begin{eqnarray}
\dot\gamma_m \cosh[H_m/\ell(\mathcal{Y}_m)] & = & \dot\gamma_+\exp[H_m/\ell(\mathcal{Y}_b)]
\nonumber \\
& + & \dot\gamma_-\exp[-H_m/\ell(\mathcal{Y}_b)]
\\
\frac{\dot\gamma_m}{\ell(\mathcal{Y}_m)}\sinh[H_m/\ell(\mathcal{Y}_m)] & = & \frac{\dot\gamma_+}{\ell(\mathcal{Y}_b)}\exp[H_m/\ell(\mathcal{Y}_b)],
\nonumber \\
& + & \frac{\dot\gamma_-}{\ell(\mathcal{Y}_b)}\exp[-H_m/\ell(\mathcal{Y}_b)].
\qquad
\end{eqnarray}
In addition, the shear rate at the interface $z=H_b$  can be expressed as:
\begin{equation}
 \dot\gamma_b=\dot\gamma_+\exp\left[\frac{H_m+H_b}{\ell(\mathcal{Y}_b)}\right]+\dot\gamma_-\exp\left[-\frac{H_m+H_b}{\ell(\mathcal{Y}_b)}\right]
 \end{equation}
We can now express $\dot\gamma_+$ and $\dot\gamma_-$ in function of $\dot\gamma_m$ and then get:
\begin{eqnarray}
\frac{{\dot \gamma}_{b}}{|{\dot \gamma}_{m}|} & = & \frac{1}{2\ell(\mathcal{Y}_m)} 
\Big[ \, \ell(\mathcal{Y}_m) \cosh{\left[ {H_m}/ \ell(\mathcal{Y}_m)\right]} e^{ H_b/ \ell(\mathcal{Y}_b)}
\nonumber \\
&+& \ell(\mathcal{Y}_b) \sinh{[ {H_m}/ \ell(\mathcal{Y}_m)]} e^{ H_b/ \ell(\mathcal{Y}_b)}
\nonumber \\
&+& \ell(\mathcal{Y}_m \cosh{\left[ {H_m}/ \ell(\mathcal{Y}_m)\right]} e^{ - H_b/ \ell(\mathcal{Y}_b)}
\nonumber \\
&-& \ell(\mathcal{Y}_b)  \sinh{[ {H_m}/ \ell(\mathcal{Y}_m)]} e^{ - H_b/ \ell(\mathcal{Y}_b)} \, \Big]
\end{eqnarray}
In the limit $H_b\gg\ell(\mathcal{Y}_b)$, the above expression simplifies into:
\begin{equation}
\frac{{\dot \gamma}_{b}}{|{\dot \gamma}_{m}|} = \frac{1}{2}
e^{\frac{H_b}{\ell(\mathcal{Y}_b)}} \!
\left[\cosh\!\left( \frac{H_{m}}{\ell(\mathcal{Y}_{m})}\right) \! +  \frac{\ell(\mathcal{Y}_b)}{\ell(\mathcal{Y}_{m})}\sinh\!\left( \frac{H_{m}}{\ell(\mathcal{Y}_{m})}\right) \! \right] \!. \qquad
\label{gammadotsSI}
\end{equation}
%

%----------------------------------------------------------------------------------------------------------------
\section{Computation of $\dot\Gamma$}
We introduce here the cross-grained field $\dot\Gamma({\vec{r},t})$, which evaluates the local rate of deformation at position $\vec{r}$ and time $t$. We take for cross-graining function a Gaussian of width $\delta$. We define $\dot\Gamma$ at the location $\vec{r}_i=(x_i,z_i)$ of the grain $i$ as
\begin{eqnarray}
\dot\Gamma(\vec{r}_i,t)=\frac{\sum \limits_{j=1}^N(u_i-u_j)(z_i-z_j)\exp\left(-\frac{||\Delta\vec{r}||^2}{2\delta^2}\right)}{\sum \limits_{j=1}^N(z_i-z_j)^2\exp\left(-\frac{||\Delta\vec{r}||^2}{2\delta^2}\right)} \, ,
\label{defGammadot}
\end{eqnarray}
where $u_i$ (resp. $u_j$) is the horizontal velocity of the grain $i$ (resp. grain $j$) at time $t$, and $||\Delta\vec{r}||=\sqrt{(z_i-z_j)^2+(x_i-x_j)^2}$ denotes the distance between the grains $i$ and $j$ at time $t$. With this definition, the space average of $\dot\Gamma$ in the centre of the shear cell gives the average shear rate $\dot\gamma$ extracted by fitting the velocity profile. In all the results presented in this Letter, $\delta$ has been set to $d$.

%___________________________________________________________________________

\end{document}